\documentclass[a4paper,fleqn,usenatbib]{mnras}

\usepackage{newtxtext,newtxmath}
\usepackage[T1]{fontenc}
\usepackage{ae,aecompl}

\usepackage{graphicx}	% Including figure files
\usepackage{amsmath}	% Advanced maths commands
\usepackage{amssymb}	% Extra maths symbols

\title[Ozone Measurements with Meteors: A Revisit]{Ozone Measurements with Meteors: A Revisit}

\author[Ye \& Han]{
Quan-Zhi Ye,$^{1,2,3}$\thanks{E-mail: qye@caltech.edu}
and Summer Xia Han$^{4,5}$
\\
$^{1}$Department of Physics and Astronomy, The University of Western Ontario, London, ON N6A 3K7, Canada\\
$^{2}$Astronomy Department, California Institute of Technology, Pasadena, CA 91125, U.S.A.\\
$^{3}$Infrared Processing and Analysis Center, California Institute of Technology, Pasadena, CA 91125, U.S.A.\\
$^{4}$Department of Biological Sciences, National University of Singapore, 117543 Singapore\\
$^{5}$Center for Craniofacial Molecular Biology, University of Southern California, Los Angeles, CA 90033, U.S.A.
}

\date{Accepted XXX. Received YYY; in original form ZZZ}

\pubyear{2015}

\begin{document}
\label{firstpage}
\pagerange{\pageref{firstpage}--\pageref{lastpage}}
\maketitle

\begin{abstract}
Understanding the role of ozone in the Mesosphere/Lower Thermosphere (MLT) region is essential for understanding the atmospheric processes in the upper atmosphere. Earlier studies have shown that it is possible to use overdense meteor trails to measure ozone concentration in the meteor region. Here we revisit this topic by comparing a compilation of radar observations to satellite measurements. We observe a modest agreement between the values derived from these two methods, which confirm the usefulness of the meteor trail technique for measuring ozone content at certain heights in the MLT region. Future simultaneous measurements will help quantifying the performance of this technique.
\end{abstract}


%%%%%%%%%%%%%%%%%%%%%%%%%%%%%%%%%%%%%%%%%%%%%%%%%%

%%%%%%%%%%%%%%%%% BODY OF PAPER %%%%%%%%%%%%%%%%%%

\section{Introduction}

A \textit{meteor} is the visible streak of light produced by an interplanetary dust particle (a \textit{meteoroid}) at the entry of the Earth's atmosphere. Meteoroids are directly linked to the primitive materials in the early solar system: they are either leftovers of the planet formation era in the early solar system, or are linked to primitive bodies (asteroids and comets), and therefore they attract intense interests in the astrophysics and planetary science communities.

The atmospheric science community, on the other hand, has used the meteor phenomenon as a tool to study the atmospheric properties and dynamics of the so-called ``meteor region'' within the Mesosphere/Lower Thermosphere (MLT) since the 1950s. This is a region 70--120~km above Earth's surface is generally difficult to study with in situ techniques. Radar meteor techniques, in particular, have been used to infer atmospheric properties such as  high-altitude temperature and winds \citep[e.g.][and many others]{fraser1965measurement,RDS:RDS3670,hocking1999temperatures,younger2008modeling,younger2014effects}, planetary-scale features \citep{cevolani1991strato,2001ESASP.495..387B}, and ozone concentration \citep[e.g.][]{jon90,jon95,haj99,cev09}. Determination of the ozone concentration uses the fact that the duration of overdense meteor trails are concurrently limited by ambipolar diffusion at high altitudes and oxidation due to the presence of ozone at lower heights as first recognized by \citet{1972MNRAS.159..203B}.

Ozone plays a major role at high altitudes (50--100~km) in the chemistry of the upper atmosphere \citep[c.f.][and references therein]{allen1984vertical}, yet measurements of ozone in the MLT remain relatively scarce. Stellar occultation technique has been routinely used to measure mesospheric ozone content since 1970s \citep[e.g.][]{1973P&SS...21..273H,1996GeoRL..23.2317B,JGRD:JGRD10123,kyr06}, but only a handful of these surveys extend their measurements to MLT and even if they do, the uncertainty in the MLT region is considerably higher. The 1.27 $\mu$m airglow emission is another technique that has been used to derive MLT ozone content \citep[e.g.][]{sic93,JGRA:JGRA14252}, but it depends on the assumption of a steady state photochemical model which does not always holds in the MLT region \citep{JGRD:JGRD13799}. An alternative technique will be helpful in providing an independent measurement of the ozone content in the MLT region.

\citet{jon90} were the first to suggest a method whereby the distribution of overdense meteor echo durations could be used to estimate ozone levels at a particular height. In the absence of chemistry-limiting reactions, the cumulative distribution of echo duration should be a power-law reflecting the distribution in the masses of incoming meteoroids, as the duration of an overdense echo subject to ambipolar diffusion is directly related to the mass of the original meteoroid \citep[e.g.][]{cep98}. Chemical reactions remove electrons from the trail and this process exceeds diffusion at longer durations. Therefore, the resulting cumulative distribution shows a break or inflection in the original power-law. The location of this inflection is a direct estimate of the ozone concentration at the height appropriate to the knee in the distribution.

Several dedicated, long-term meteor radar systems, such as the Advanced Meteor Orbit Radar in 1990--1999 \citep{baggaley1995radar}, the Canadian Meteor Orbit Radar (CMOR) from 2002 onward \citep{jones2005canadian}, and meteor radars at Antarctica from 2006 onward \citep{2008AdSpR..42..143H,2009MNRAS.398..350Y}, each have collected orders of magnitude more data than their precursors. However, unlike temperature and wind measurements, which rely on underdense meteor trails that are relatively easy to identify with automatic algorithms, determination of ozone concentration relies on the precise measurement of the duration of overdense meteor trails, a process that is difficult to automate. Therefore, ozone study takes little advantage from the dramatic increase of meteor data.

Here, we revisit this topic combining results of a number of earlier studies, including several CMOR datasets that are recently analyzed and published \citep[][and others]{ye2013radar,ye2013unexpected}. Our goal is to compare the resulting ozone estimates with more recent satellite estimates of ozone with overlapping temporal coverage, updated reaction rate coefficients, and critically review the utility of this technique for providing ozone concentrations at specific heights.

\section{Theory: A Revisit}

The mass distribution index is defined such that the number of meteoroids in the mass interval $(m,m+dm)$ follows $m^{-s}$ \citep{mck61,Grun1985b}. In most cases, the assumption of a power-law distribution of meteoroid number as a function of mass holds for a wide mass range, although in practice, most streams are found to have a unique $s$.

It is difficult to directly measure the mass of an individual meteoroid, but following classic radar meteor theory, we expect the mass of a meteoroid to be linearly proportional to the (peak) electron line density $q$ of the trail formed (for underdense meteor trails) or the \textit{height-corrected} \citep{1987BAICz..38...80S} duration $\tau$ (for overdense meteor trails) as seen by the radar. Therefore, the distribution of $q$ or $\tau$ can be used to constrain $s$. This can be done by simply counting the cumulative number of meteors $N$ beyond a certain value of $q$ or $\tau$ and measure the slope of the linear portion to derive $s$ \citep{mci68}:

\begin{equation}
N \propto q^{1-s}
\end{equation}

\noindent for underdense trails, or

\begin{equation}
N \propto \tau^{3(1-s)/4}
\label{ovd-mass}
\end{equation}

\noindent for overdense trails.

Equation~\ref{ovd-mass} provides the theoretical duration of an overdense meteor trail when the decay process is dominated by ambipolar diffusion. Larger/slower meteoroids can survive to lower heights (below $\sim 100$~km), where the duration can be limited by the reaction rate of a two-body attachment process between meteoric electrons, called the \textit{chemistry-limited~regime}. Diffusion of meteor trails above $\sim 100$~km are predominately controlled by ambipolar process and is not considered susceptible to chemistry-limited regime.

The duration in the chemistry-limited regime is expressed as follow \citep[][\S 1]{jon90}, although it has been recognized that the change of power beam collecting area can alter its slope \citep{Pecina1984b, Pecinova2005}:

\begin{equation}
N \propto \tau^{9(1-s)/2}
\end{equation}

\begin{figure}
\noindent\includegraphics[width=0.5\textwidth]{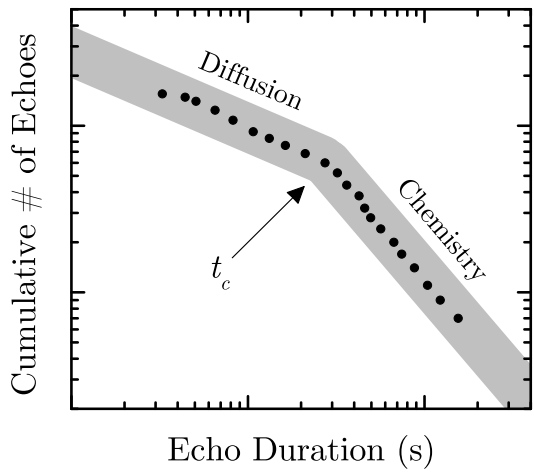}
\caption{The determination of the characteristic duration, $t_c$, presented as a turnover point from diffusion-limited regime to chemistry-limited region. This illustrative figure does not involve real data.}
\label{fig-example}
\end{figure}

However, the derived $s$ values showed large discrepancies with values derived from independent techniques (e.g. optical measurements), possibly due to the non-negligible contributions from other factors, such as gravity waves or charged dust particles \citep[e.g.][]{kel04}. Nonetheless, the transition between diffusion-limited regime and chemistry-limited regime is marked by the ``characteristic'' duration, $t_c$ (Figure~\ref{fig-example}). Trails with duration greater than $t_c$ are predominantly in the chemistry-limited regime and vice versa.

\citet{bag74} attribute the cause of chemistry-limited regime to rapid dissociative recombination between meteoric ions and ozone molecules, which was later supported by observational evidence \citep{jon90}. The key reactions removing electrons from the trail are:

\begin{equation}
M^+ + \mathrm{O_3} \rightarrow M\mathrm{O}^+ + \mathrm{O_2}
\label{reaction1}
\end{equation}

\begin{equation}
M\mathrm{O}^+ + e \rightarrow M + \mathrm{O}
\label{reaction2}
\end{equation}

\noindent where $M^+$ is the meteoric ion. The reaction constant of Reaction~\ref{reaction1} is 2--3 orders of magnitude slower than Reaction~\ref{reaction2} \citep{wha11,plane2012new}, which limits the de-ionization process. We should note that this reaction is only dominant above $\sim 88$~km \citep{fer68,rowe1981flowing,wha11}, hence limiting the heights that the meteor trail technique can be applied.

Readers may also wonder about the reaction

\begin{equation}
M\mathrm{O}^+ + O \rightarrow M^+ + \mathrm{O}_2
\label{reaction2a}
\end{equation}

\noindent which dominates over Reaction~\ref{reaction2} in the background MLT. However, it should be noted that in \textit{overdense} meteor trails, the electron density \citep[$\sim10^{20}~\mathrm{m^{-3}}$, e.g.][]{1999A&A...341..634F} is much higher than the ambient atomic oxygen density \citep[$\sim10^{18}~\mathrm{m^{-3}}$, e.g.][]{amt-8-1021-2015} and therefore, Reaction~\ref{reaction2} will dominate over Reaction~\ref{reaction2a} in the overdense meteor trails.

Taking the reaction constant of Reaction~\ref{reaction1} to be $k$, the ozone concentration can be calculated by

\begin{equation}
[\mathrm{O_3}] = \frac{1}{k \cdot t_c}
\label{o3_compute}
\end{equation}

\noindent as applicable to the knee height of the meteors, namely the ``knee'' height where $t_c$ applies.

For simplicity, Mg$^+$ has been used as the representative species for major meteoric ions \citep[e.g.][]{cervera2000comparison,younger2014effects}. However, it is known that other species that participate the oxidation process, such as Si$^+$ and Fe$^+$, are also major ion species in meteoroids\citep{bag74}. Although oxidation with Mg$^+$ ion is slightly more efficient than that of Si$^+$ and Fe$^+$ (Table~\ref{tbl-abundance}), some meteoroid streams (such as $\eta$-Aquariids and Orionids) are known to be Si-rich \citep{jes88}. If we take the whole process as a simple stoichiometric reaction, we have

\begin{equation}
k = \sum_{i=1}^{N} w_i k_i
\label{k_compute}
\end{equation}

\noindent where $w_i$ and $k_i$ denote the weight and reaction constant, respectively, of the $i$-th chemical species, which is either Mg$^+$, Si$^+$ or Fe$^+$ in our work. The weight is derived using relative abundance and fraction of ionization reported in earlier papers \citep{1997MNRAS.288..995J,von08} for each meteoroid stream. We follow \citet{bag74}'s suggestion that Mg$^+$, Si$^+$ and Fe$^+$ make up 93\% of meteoric ions and assume Mg$^+$, Si$^+$ and Fe$^+$ are fully responsible for oxidation process (the remaining 7\% is attributed to Na$^+$ which does not take part in the oxidation process).

\section{Data Collection and Results}

\begin{figure}
	\noindent\includegraphics[width=\textwidth]{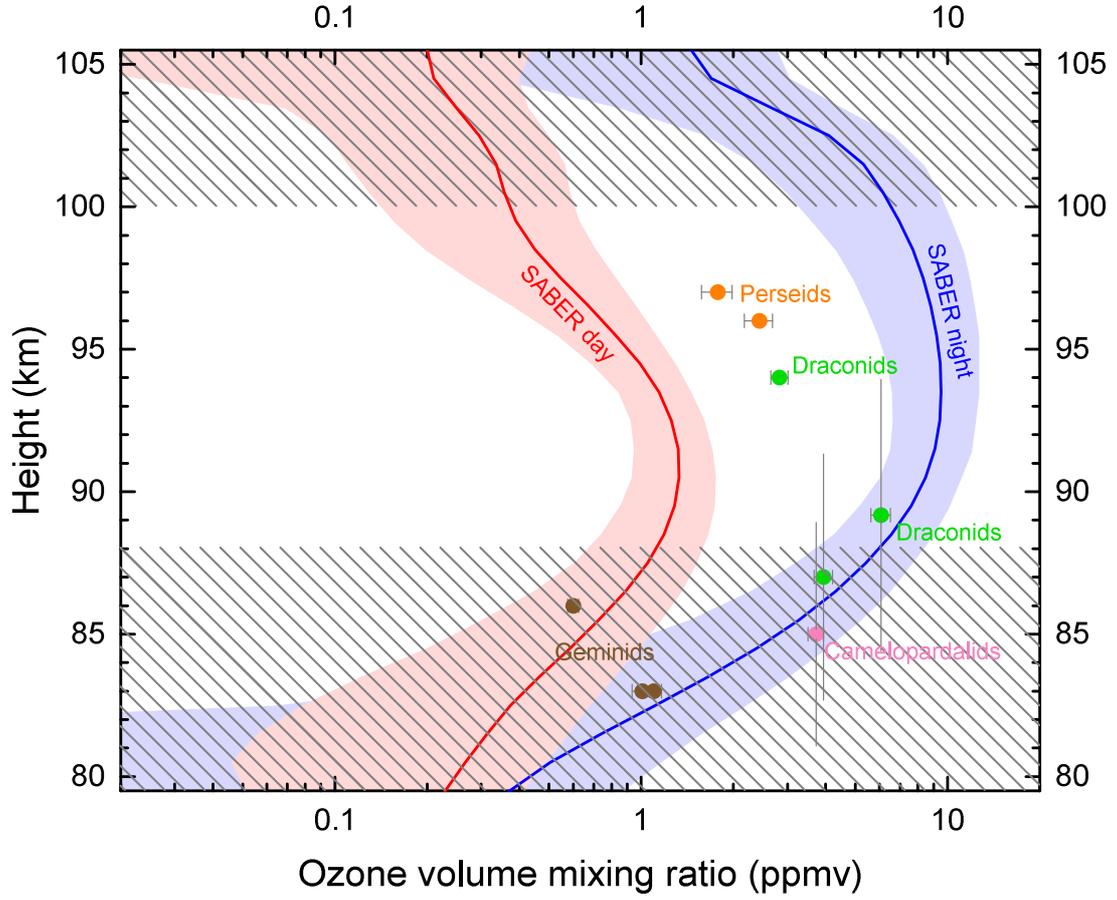}
	\caption{Ozone mixing ratio (in part per million volumes) derived from meteor trails and SABER. Uncertainty bars indicate the standard deviation of the SABER and meteor-derived data. Shaded area indicates region that the meteor technique is theoretically invalid.}
	\label{fig-ozone}
\end{figure}

We compiled a total of 9 reported height-corrected characteristic times $t_c$, including data from nine meteoroid streams recorded by five radar systems with observations dating back to 1957. Unfortunately, most of the works do not report the uncertainty of their data. Moreover, the process of fitting is not always clearly documented, making the results difficult to reproduce. Hence, we re-measure and re-fit all dataset using a linear piecewise function which will also give us the fitting error. All re-measured $t_c$ agree the original values within $\sim30\%$. We then converted $t_c$ to ozone concentration using Equation~\ref{o3_compute} and Equation~\ref{k_compute} with weights and combined reaction constants for the corresponding meteoroid stream. The results are shown in Table~\ref{tbl-results} and Figure~\ref{fig-ozone}.

With regard to the determination of the knee height, or the knee height of the meteors, there are three approaches:

\begin{enumerate}
  \item Interferometric observations, which the specular height of every observed meteor is measured directly by the radar. The knee height can therefore be taken as the mean inflection height of the meteors in the sample.
  \item Simultaneous radar-visual observations \citep{jon90}, in which meteors are simultaneously observed by radar and a set of visual observers, the knee height is derived using the relation between meteor luminosity and its radio duration \citep[][\S8-13]{mck61}.
  \item Photographic observations, which applies to the radar systems with no interferometer or simultaneous visual observations. The average meteor luminosity is taken from other photographic or video meteor surveys \citep[e.g.][]{jac67,bro02}, allowing knee height to be derived using luminosity-duration relation. \citet{jon95} shows that the uncertainty of this method is $\sim 3$~km.
\end{enumerate}

CMOR is equipped with an interferometer, Springhill is accompanied by visual observers. The other two radar systems (Ond\v{r}ejov and Kharkov) use photographic mean determined by \citet{jac67} and \citet[][for Leonids]{bro02} to derive the knee height.

We then compared our results to those derived from SABER (Sounding of the Atmosphere using Broadband Emission Radiometry) instrument on the TIMED (Thermosphere, Ionosphere, Mesosphere, Energetics and Dynamics) satellite which measures ozone profile using star occultation by Earth's atmosphere. The 9.6~$\mu$m ozone data obtained by SABER in March, June, September and December in 2002--2013 within $10^{\circ}$ longitude and $5^{\circ}$ latitude around CMOR site are extracted from the level 2 V2.0 datasets as available through the SABER website (\url{http://saber.gats-inc.com/index.php}).

As shown in Figure~\ref{fig-ozone}, we observe a modest agreement between the meteor-derived values and SABER measurements within the region which the meteor trail method is valid (88--100~km): all meteor-derived data points are within the range of diurnal variation of ozone as measured by SABER. This confirms earlier studies that meteor trail technique is effective. Beyond the 88--100~km region, the meteor-derived value still show some trend of agreement with the SABER values, but it is difficult to judge how far this agreement extend beyond the valid region and whether it can be trusted.

However, the lack of time information of the meteor data prevent us to compare the meteor data and SABER data more directly (i.e. data taken at the same time). This prompts us to focus on data from meteor outbursts only. There are three meteor outburst events in our dataset: the Draconid outbursts in 2011 Oct 8 (16--20~h UT) and 2012 Oct 8 (15--19~h UT) \citep{ye2013radar,ye2013unexpected}, and the Camelopardalid outburst in 2014 May 24 (4--12~h UT) \citep{ye2013will}. Meteor outburst is a phenomenon in which meteor rate from a particular meteoroid stream increases significantly, usually just in a few hours. The increase of meteor rate helps to improve the statistics, while the short duration helps by reducing the potential blur caused by temporal variation of ozone content. However, this introduces another difficulty as SABER do not always have observations in the desired time period. To increase the statistics at the SABER end, we include SABER data taken in the same hours within five days from the outburst dates. Still, the SABER data taken at both Draconid dates were 4--6~h too early for the outbursts, which would translate to an ozone level about 2 times lower compared to the ozone level at the outburst times which were both at local noon \citep[][Figure 8b]{hua08}.

\begin{figure}
	\noindent\includegraphics[width=\textwidth]{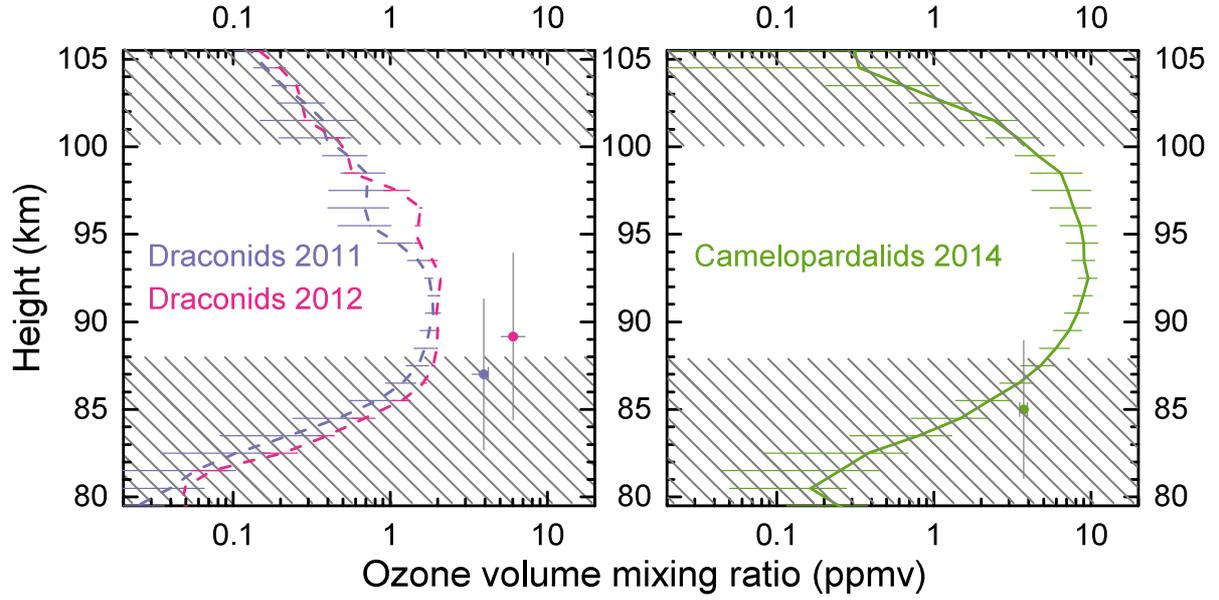}
	\caption{Semi-simultaneous ozone measurements by CMOR (circles) and SABER during Draconids outbursts in 2011 and 2012 (left) and Camelopardalids outburst in 2014 (right). CMOR measurements are appropriated to 2011 Oct 8 at 16--20~h UT and 2012 Oct 8 at 15--19~h UT for the Draconids, and 2014 May 24 at 4--12~h UT for the Camelopardalids. SABER observations are conducted at similar hours within five days from the outburst dates. SABER profiles appropriated for the Draconid outbursts are plotted in dashed lines, as the observations are 4--6~h too early for the meteor-derived value. Uncertainty bars for SABER profile and CMOR data indicates the standard deviation of the data. Shaded area indicates region that the meteor technique is theoretically invalid.}
	\label{fig-cmor}
\end{figure}

As shown in Figure~\ref{fig-cmor}, the results are unfortunately suffered from low statistics, as well as suffered from the fact that most data points are beyond the valid zone for the meteor trail technique. Future campaigns aiming at collecting more simultaneous measurements will be useful to evaluate the performance of the meteor trail technique.

We also note that the compositional difference among the streams seems to have minor impact on the derivation of ozone. Our sample covers a wide range of compositions, from Mg-rich (such as Draconids) to Si-rich (such as Perseids), but the resulting combined rate constants are within $\sim 5\%$ from each other. These showers are major annual showers that can be considered to be representative to shower meteors. Despite Mg$^+$ having a higher rate constant, the ionization fraction of Mg is always lower than that of Si and Fe within the entire speed range, thus the Mg$^+$ reaction will not dominate under any circumstance.

\section{Summary}

We revisited the technique of using overdense meteor trails to measure ozone content in the MLT region. The technique was examined by comparing data derived from radar meteor observations to the ones derived from satellite observations. We observe a modest agreement between the two, confirming the results reported by earlier studies. However, the lack of simultaneous measurements made by different techniques prevent further evaluation about the performance of the meteor technique. Future campaigns focused in collecting more simultaneous measurements of the meteor technique and other techniques would be useful to resolve this issue.

\section*{Acknowledgments}
We thank several anonymous referees for their comments, as well as Peter Brown, Chen Hai-Sheng, Wang Hong-Lei, and Jia Shi-Guo for the discussion. We also thank the SABER team for their work in preparing the SABER dataset (available through \url{http://saber.gats-inc.com/index.php}) and making it readily available.

\bibliographystyle{mnras}
\bibliography{ms}

\begin{thebibliography}{}
\makeatletter
\relax
\def\mn@urlcharsother{\let\do\@makeother \do\$\do\&\do\#\do\^\do\_\do\%\do\~}
\def\mn@doi{\begingroup\mn@urlcharsother \@ifnextchar [ {\mn@doi@}
  {\mn@doi@[]}}
\def\mn@doi@[#1]#2{\def\@tempa{#1}\ifx\@tempa\@empty \href
  {http://dx.doi.org/#2} {doi:#2}\else \href {http://dx.doi.org/#2} {#1}\fi
  \endgroup}
\def\mn@eprint#1#2{\mn@eprint@#1:#2::\@nil}
\def\mn@eprint@arXiv#1{\href {http://arxiv.org/abs/#1} {{\tt arXiv:#1}}}
\def\mn@eprint@dblp#1{\href {http://dblp.uni-trier.de/rec/bibtex/#1.xml}
  {dblp:#1}}
\def\mn@eprint@#1:#2:#3:#4\@nil{\def\@tempa {#1}\def\@tempb {#2}\def\@tempc
  {#3}\ifx \@tempc \@empty \let \@tempc \@tempb \let \@tempb \@tempa \fi \ifx
  \@tempb \@empty \def\@tempb {arXiv}\fi \@ifundefined
  {mn@eprint@\@tempb}{\@tempb:\@tempc}{\expandafter \expandafter \csname
  mn@eprint@\@tempb\endcsname \expandafter{\@tempc}}}

\bibitem[\protect\citeauthoryear{Allen, Lunine  \& Yung}{Allen
  et~al.}{1984}]{allen1984vertical}
Allen M.,  Lunine J.~I.,   Yung Y.~L.,  1984, Journal of Geophysical Research:
  Atmospheres (1984--2012), 89, 4841

\bibitem[\protect\citeauthoryear{{Anders} \& {Grevesse}}{{Anders} \&
  {Grevesse}}{1989}]{and89}
{Anders} E.,  {Grevesse} N.,  1989, \mn@doi [Geochimica et Cosmochimica Acta]
  {10.1016/0016-7037(89)90286-X}, \href
  {http://adsabs.harvard.edu/abs/1989GeCoA..53..197A} {53, 197}

\bibitem[\protect\citeauthoryear{{Baggaley}}{{Baggaley}}{1972}]{1972MNRAS.159..203B}
{Baggaley} W.~J.,  1972, Monthly Notices of the Royal Astronomical Society,
  \href {http://adsabs.harvard.edu/abs/1972MNRAS.159..203B} {159, 203}

\bibitem[\protect\citeauthoryear{Baggaley}{Baggaley}{1995}]{baggaley1995radar}
Baggaley W.,  1995, Earth, Moon, and Planets, 68, 127

\bibitem[\protect\citeauthoryear{{Baggaley} \& {Cummack}}{{Baggaley} \&
  {Cummack}}{1974}]{bag74}
{Baggaley} W.~J.,  {Cummack} C.~H.,  1974, Journal of Atmospheric and
  Terrestrial Physics, \href
  {http://adsabs.harvard.edu/abs/1974JATP...36.1759B} {36, 1759}

\bibitem[\protect\citeauthoryear{{Baggaley}, {Marsh}, {Bennett}  \&
  {Galligan}}{{Baggaley} et~al.}{2001}]{2001ESASP.495..387B}
{Baggaley} W.~J.,  {Marsh} S.~H.,  {Bennett} R.~G.~T.,   {Galligan} D.~P.,
  2001, in {Warmbein} B.,  ed.,  ESA Special Publication Vol. 495, Meteoroids
  2001 Conference. pp 387--391

\bibitem[\protect\citeauthoryear{{Bevilacqua} et~al.,}{{Bevilacqua}
  et~al.}{1996}]{1996GeoRL..23.2317B}
{Bevilacqua} R.~M.,  et~al., 1996, \mn@doi [Geophysical Research Letters]
  {10.1029/96GL01119}, \href
  {http://adsabs.harvard.edu/abs/1996GeoRL..23.2317B} {23, 2317}

\bibitem[\protect\citeauthoryear{Borovi{\v{c}}ka}{Borovi{\v{c}}ka}{2005}]{borovivcka2005elemental}
Borovi{\v{c}}ka J.,  2005, in , Modern Meteor Science An Interdisciplinary
  View.
Dordrecht, Netherlands: Springer, pp 245--253

\bibitem[\protect\citeauthoryear{{Borovi{\v c}ka}, {Spurn{\'y}}  \&
  {Koten}}{{Borovi{\v c}ka} et~al.}{2007}]{bor07}
{Borovi{\v c}ka} J.,  {Spurn{\'y}} P.,   {Koten} P.,  2007, \mn@doi [Astronomy
  \& Astrophysics] {10.1051/0004-6361:20078131}, \href
  {http://adsabs.harvard.edu/abs/2007A%26A...473..661B} {473, 661}

\bibitem[\protect\citeauthoryear{{Brown}, {Campbell}, {Hawkes}, {Theijsmeijer}
  \& {Jones}}{{Brown} et~al.}{2002}]{bro02}
{Brown} P.,  {Campbell} M.~D.,  {Hawkes} R.~L.,  {Theijsmeijer} C.,   {Jones}
  J.,  2002, \mn@doi [Planetary and Space Science]
  {10.1016/S0032-0633(01)00112-X}, \href
  {http://adsabs.harvard.edu/abs/2002P%26SS...50...45B} {50, 45}

\bibitem[\protect\citeauthoryear{{Ceplecha}, {Borovi{\v c}ka}, {Elford},
  {Revelle}, {Hawkes}, {Porub{\v c}an}  \& {{\v S}imek}}{{Ceplecha}
  et~al.}{1998}]{cep98}
{Ceplecha} Z.,  {Borovi{\v c}ka} J.,  {Elford} W.~G.,  {Revelle} D.~O.,
  {Hawkes} R.~L.,  {Porub{\v c}an} V.,   {{\v S}imek} M.,  1998, \mn@doi [Space
  Science Reviews] {10.1023/A:1005069928850}, \href
  {http://adsabs.harvard.edu/abs/1998SSRv...84..327C} {84, 327}

\bibitem[\protect\citeauthoryear{Cervera \& Reid}{Cervera \&
  Reid}{1995}]{RDS:RDS3670}
Cervera M.~A.,  Reid I.~M.,  1995, \mn@doi [Radio Science] {10.1029/95RS00644},
  30, 1245

\bibitem[\protect\citeauthoryear{Cervera \& Reid}{Cervera \&
  Reid}{2000}]{cervera2000comparison}
Cervera M.~A.,  Reid I.~M.,  2000, Radio Science, 35, 833

\bibitem[\protect\citeauthoryear{Cevolani}{Cevolani}{1991}]{cevolani1991strato}
Cevolani G.,  1991, Geophysical research letters, 18, 1987

\bibitem[\protect\citeauthoryear{Cevolani \& Pupillo}{Cevolani \&
  Pupillo}{2003}]{cev09}
Cevolani G.,  Pupillo G.,  2003, Annals of Geophysics, 46

\bibitem[\protect\citeauthoryear{Ferguson \& Fehsenfeld}{Ferguson \&
  Fehsenfeld}{1968}]{fer68}
Ferguson E.,  Fehsenfeld F.,  1968, \mn@doi [J. Geophys. Res.]
  {10.1029/JA073i019p06215}, 73, 6215

\bibitem[\protect\citeauthoryear{{Foschini}}{{Foschini}}{1999}]{1999A&A...341..634F}
{Foschini} L.,  1999, Astronomy \& Astrophysics, \href
  {http://adsabs.harvard.edu/abs/1999A%26A...341..634F} {341, 634}

\bibitem[\protect\citeauthoryear{Fraser}{Fraser}{1965}]{fraser1965measurement}
Fraser G.,  1965, Journal of Atmospheric Sciences, 22, 217

\bibitem[\protect\citeauthoryear{Gomez~Martin \& Plane}{Gomez~Martin \&
  Plane}{2011}]{gom11}
Gomez~Martin J.~C.,  Plane J. M.~C.,  2011, \mn@doi [Phys. Chem. Chem. Phys.]
  {10.1039/C0CP01380C}, 13, 3764

\bibitem[\protect\citeauthoryear{{Grun}, {Zook}, {Fechtig}  \& {Giese}}{{Grun}
  et~al.}{1985}]{Grun1985b}
{Grun} E.,  {Zook} H.~A.,  {Fechtig} H.,   {Giese} R.~H.,  1985, \mn@doi
  [Icarus] {10.1016/0019-1035(85)90121-6}, \href
  {http://ads.ari.uni-heidelberg.de/abs/1985Icar...62..244G} {62, 244}

\bibitem[\protect\citeauthoryear{Gumbel, Murtagh, Espy, Witt  \&
  Schmidlin}{Gumbel et~al.}{1998}]{JGRA:JGRA14252}
Gumbel J.,  Murtagh D.~P.,  Espy P.~J.,  Witt G.,   Schmidlin F.~J.,  1998,
  \mn@doi [Journal of Geophysical Research: Space Physics] {10.1029/98JA02155},
  103, 23399

\bibitem[\protect\citeauthoryear{{Hajduk}, {Hajdukov{\'a}}, {Porub{\v c}an},
  {Cevolani}  \& {Grassi}}{{Hajduk} et~al.}{1999}]{haj99}
{Hajduk} A.,  {Hajdukov{\'a}} M.,  {Porub{\v c}an} V.,  {Cevolani} G.,
  {Grassi} G.,  1999, in {Baggaley} W.~J.,  {Porubcan} V.,  eds, Proceedings of
  the International Conference held at Tatranska Lomnica, Slovakia, August
  17-21, 1998. p.~91

\bibitem[\protect\citeauthoryear{{Hays} \& {Roble}}{{Hays} \&
  {Roble}}{1973}]{1973P&SS...21..273H}
{Hays} P.~B.,  {Roble} R.~G.,  1973, \mn@doi [Planetary and Space Science]
  {10.1016/0032-0633(73)90011-1}, \href
  {http://adsabs.harvard.edu/abs/1973P%26SS...21..273H} {21, 273}

\bibitem[\protect\citeauthoryear{Hocking}{Hocking}{1999}]{hocking1999temperatures}
Hocking W.,  1999, Geophysical Research Letters, 26, 3297

\bibitem[\protect\citeauthoryear{{Holdsworth}, {Murphy}, {Reid}  \&
  {Morris}}{{Holdsworth} et~al.}{2008}]{2008AdSpR..42..143H}
{Holdsworth} D.~A.,  {Murphy} D.~J.,  {Reid} I.~M.,   {Morris} R.~J.,  2008,
  \mn@doi [Advances in Space Research] {10.1016/j.asr.2007.02.037}, \href
  {http://adsabs.harvard.edu/abs/2008AdSpR..42..143H} {42, 143}

\bibitem[\protect\citeauthoryear{{Huang}, {Mayr}, {Russell}, {Mlynczak}  \&
  {Reber}}{{Huang} et~al.}{2008}]{hua08}
{Huang} F.~T.,  {Mayr} H.~G.,  {Russell} J.~M.,  {Mlynczak} M.~G.,   {Reber}
  C.~A.,  2008, \mn@doi [Journal of Geophysical Research (Space Physics)]
  {10.1029/2007JA012739}, \href
  {http://adsabs.harvard.edu/abs/2008JGRA..11304307H} {113, 4307}

\bibitem[\protect\citeauthoryear{{Jacchia}, {Verniani}  \& {Briggs}}{{Jacchia}
  et~al.}{1967}]{jac67}
{Jacchia} L.,  {Verniani} F.,   {Briggs} R.~E.,  1967, Smithsonian
  Contributions to Astrophysics, \href
  {http://adsabs.harvard.edu/abs/1967SCoA...10....1J} {10, 1}

\bibitem[\protect\citeauthoryear{{Jessberger}, {Christoforidis}  \&
  {Kissel}}{{Jessberger} et~al.}{1988}]{jes88}
{Jessberger} E.~K.,  {Christoforidis} A.,   {Kissel} J.,  1988, \mn@doi
  [Nature] {10.1038/332691a0}, \href
  {http://adsabs.harvard.edu/abs/1988Natur.332..691J} {332, 691}

\bibitem[\protect\citeauthoryear{{Jones}}{{Jones}}{1997}]{1997MNRAS.288..995J}
{Jones} W.,  1997, Monthly Notices of the Royal Astronomical Society, \href
  {http://adsabs.harvard.edu/abs/1997MNRAS.288..995J} {288, 995}

\bibitem[\protect\citeauthoryear{{Jones} \& {Simek}}{{Jones} \&
  {Simek}}{1995}]{jon95}
{Jones} J.,  {Simek} M.,  1995, \mn@doi [Earth Moon and Planets]
  {10.1007/BF00671524}, \href
  {http://adsabs.harvard.edu/abs/1995EM%26P...68..329J} {68, 329}

\bibitem[\protect\citeauthoryear{{Jones}, {McIntosh}  \& {Simek}}{{Jones}
  et~al.}{1990}]{jon90}
{Jones} J.,  {McIntosh} B.~A.,   {Simek} M.,  1990, Journal of Atmospheric and
  Terrestrial Physics, \href
  {http://adsabs.harvard.edu/abs/1990JATP...52..253J} {52, 253}

\bibitem[\protect\citeauthoryear{Jones, Brown, Ellis, Webster, Campbell-Brown,
  Krzemenski  \& Weryk}{Jones et~al.}{2005}]{jones2005canadian}
Jones J.,  Brown P.,  Ellis K.,  Webster A.,  Campbell-Brown M.,  Krzemenski
  Z.,   Weryk R.,  2005, Planetary and Space Science, 53, 413

\bibitem[\protect\citeauthoryear{{Kasuga}, {Watanabe}  \& {Ebizuka}}{{Kasuga}
  et~al.}{2005}]{kas05}
{Kasuga} T.,  {Watanabe} J.,   {Ebizuka} N.,  2005, \mn@doi [Astronomy \&
  Astrophysics] {10.1051/0004-6361:200500142}, \href
  {http://adsabs.harvard.edu/abs/2005A%26A...438L..17K} {438, L17}

\bibitem[\protect\citeauthoryear{Kaufmann, Gusev, Grossmann, Martín-Torres,
  Marsh  \& Kutepov}{Kaufmann et~al.}{2003}]{JGRD:JGRD10123}
Kaufmann M.,  Gusev O.~A.,  Grossmann K.~U.,  Martín-Torres F.~J.,  Marsh
  D.~R.,   Kutepov A.~A.,  2003, \mn@doi [Journal of Geophysical Research:
  Atmospheres] {10.1029/2002JD002800}, 108

\bibitem[\protect\citeauthoryear{Kelley}{Kelley}{2004}]{kel04}
Kelley M.~C.,  2004, \mn@doi [Radio Sci.] {10.1029/2003RS002988}, 39, RS2015

\bibitem[\protect\citeauthoryear{Kyr{\"o}l{\"a} et~al.,}{Kyr{\"o}l{\"a}
  et~al.}{2006}]{kyr06}
Kyr{\"o}l{\"a} E.,  et~al., 2006, \mn@doi [J. Geophys. Res.]
  {10.1029/2006JD007193}, 111, D24306

\bibitem[\protect\citeauthoryear{Lednyts'kyy, von Savigny, Eichmann  \&
  Mlynczak}{Lednyts'kyy et~al.}{2015}]{amt-8-1021-2015}
Lednyts'kyy O.,  von Savigny C.,  Eichmann K.-U.,   Mlynczak M.~G.,  2015,
  \mn@doi [Atmospheric Measurement Techniques] {10.5194/amt-8-1021-2015}, 8,
  1021

\bibitem[\protect\citeauthoryear{{McIntosh}}{{McIntosh}}{1968}]{mci68}
{McIntosh} B.~A.,  1968, in {Kresak} L.,  {Millman} P.~M.,  eds,  IAU Symposium
  Vol. 33, Physics and Dynamics of Meteors. p.~343

\bibitem[\protect\citeauthoryear{{McKinley}}{{McKinley}}{1961}]{mck61}
{McKinley} D.~W.~R.,  1961, {Meteor science and engineering.}.
New York, McGraw-Hill

\bibitem[\protect\citeauthoryear{Pecina}{Pecina}{1984}]{Pecina1984b}
Pecina P.,  1984, Bulletin of the Astronomical Institutes of Czechoslovakia,
  35, 183

\bibitem[\protect\citeauthoryear{Pecinov{\'a} \& Pecina}{Pecinov{\'a} \&
  Pecina}{2005}]{Pecinova2005}
Pecinov{\'a} D.,  Pecina P.,  2005, Modern Meteor Science An Interdisciplinary
  View, pp 689--696

\bibitem[\protect\citeauthoryear{Plane \& Whalley}{Plane \&
  Whalley}{2012}]{plane2012new}
Plane J.~M.,  Whalley C.~L.,  2012, The Journal of Physical Chemistry A, 116,
  6240

\bibitem[\protect\citeauthoryear{Rollason \& Plane}{Rollason \&
  Plane}{1998}]{rol98}
Rollason R.~J.,  Plane J. M.~C.,  1998, \mn@doi [J. Chem. Soc.{,} Faraday
  Trans.] {10.1039/A805140B}, 94, 3067

\bibitem[\protect\citeauthoryear{Rowe, Fahey, Ferguson  \& Fehsenfeld}{Rowe
  et~al.}{1981}]{rowe1981flowing}
Rowe B.,  Fahey D.,  Ferguson E.,   Fehsenfeld F.,  1981, The Journal of
  Chemical Physics, 75, 3325

\bibitem[\protect\citeauthoryear{Sica \& Lowe}{Sica \& Lowe}{1993}]{sic93}
Sica R.,  Lowe R.,  1993, \mn@doi [J. Geophys. Res.] {10.1029/92JD02518}, 98,
  1051

\bibitem[\protect\citeauthoryear{{Simek}}{{Simek}}{1987}]{1987BAICz..38...80S}
{Simek} M.,  1987, Bulletin of the Astronomical Institutes of Czechoslovakia,
  \href {http://adsabs.harvard.edu/abs/1987BAICz..38...80S} {38, 80}

\bibitem[\protect\citeauthoryear{{Simek}}{{Simek}}{1994}]{sim94}
{Simek} M.,  1994, Astronomy \& Astrophysics, \href
  {http://adsabs.harvard.edu/abs/1994A%26A...284..276S} {284, 276}

\bibitem[\protect\citeauthoryear{Vondrak, Plane, Broadley  \& Janches}{Vondrak
  et~al.}{2008}]{von08}
Vondrak T.,  Plane J. M.~C.,  Broadley S.,   Janches D.,  2008, \mn@doi
  [Atmospheric Chemistry and Physics Discussions] {10.5194/acpd-8-14557-2008},
  8, 14557

\bibitem[\protect\citeauthoryear{Whalley, Martin, Wright  \& Plane}{Whalley
  et~al.}{2011}]{wha11}
Whalley C.~L.,  Martin J. C.~G.,  Wright T.~G.,   Plane J. M.~C.,  2011,
  \mn@doi [Phys. Chem. Chem. Phys.] {10.1039/C0CP02637A}, 13, 6352

\bibitem[\protect\citeauthoryear{Ye \& Wiegert}{Ye \&
  Wiegert}{2013}]{ye2013will}
Ye Q.,  Wiegert P.~A.,  2013, Monthly Notices of the Royal Astronomical
  Society, pp 3283--3287

\bibitem[\protect\citeauthoryear{Ye, Wiegert, Brown, Campbell-Brown  \&
  Weryk}{Ye et~al.}{2013a}]{ye2013unexpected}
Ye Q.,  Wiegert P.~A.,  Brown P.~G.,  Campbell-Brown M.~D.,   Weryk R.~J.,
  2013a, Monthly Notices of the Royal Astronomical Society, p. stt2178

\bibitem[\protect\citeauthoryear{Ye, Brown, Campbell-Brown  \& Weryk}{Ye
  et~al.}{2013b}]{ye2013radar}
Ye Q.,  Brown P.~G.,  Campbell-Brown M.~D.,   Weryk R.~J.,  2013b, Monthly
  Notices of the Royal Astronomical Society, 436, 675

\bibitem[\protect\citeauthoryear{{Ye}, {Hui}, {Brown}, {Campbell-Brown},
  {Pokorn{\'y}}, {Wiegert}  \& {Gao}}{{Ye} et~al.}{2016}]{2016Icar..264...48Y}
{Ye} Q.-Z.,  {Hui} M.-T.,  {Brown} P.~G.,  {Campbell-Brown} M.~D.,
  {Pokorn{\'y}} P.,  {Wiegert} P.~A.,   {Gao} X.,  2016, \mn@doi [Icarus]
  {10.1016/j.icarus.2015.09.003}, \href
  {http://adsabs.harvard.edu/abs/2016Icar..264...48Y} {264, 48}

\bibitem[\protect\citeauthoryear{Younger, Reid, Vincent  \& Holdsworth}{Younger
  et~al.}{2008}]{younger2008modeling}
Younger J.~P.,  Reid I.~M.,  Vincent R.~A.,   Holdsworth D.~A.,  2008,
  Geophysical Research Letters, 35

\bibitem[\protect\citeauthoryear{{Younger}, {Reid}, {Vincent}, {Holdsworth}  \&
  {Murphy}}{{Younger} et~al.}{2009}]{2009MNRAS.398..350Y}
{Younger} J.~P.,  {Reid} I.~M.,  {Vincent} R.~A.,  {Holdsworth} D.~A.,
  {Murphy} D.~J.,  2009, \mn@doi [Monthly Notices of the Royal Astronomical
  Society] {10.1111/j.1365-2966.2009.15142.x}, \href
  {http://adsabs.harvard.edu/abs/2009MNRAS.398..350Y} {398, 350}

\bibitem[\protect\citeauthoryear{Younger, Lee, Reid, Vincent, Kim  \&
  Murphy}{Younger et~al.}{2014}]{younger2014effects}
Younger J.,  Lee C.,  Reid I.,  Vincent R.,  Kim Y.,   Murphy D.,  2014,
  Journal of Geophysical Research: Atmospheres, 119, 10027

\bibitem[\protect\citeauthoryear{Zhu, Yee  \& Talaat}{Zhu
  et~al.}{2007}]{JGRD:JGRD13799}
Zhu X.,  Yee J.-H.,   Talaat E.~R.,  2007, \mn@doi [Journal of Geophysical
  Research: Atmospheres] {10.1029/2007JD008447}, 112

\makeatother
\end{thebibliography}

\clearpage

\clearpage

\clearpage

\clearpage

\begin{table}
\caption{Weight and combined rate constant ($k$) derived for each meteoroid stream. For abundance ratio, numbers in brackets indicate that ratio for the respective stream is unknown, therefore generalize value from carbonaceous (CI) chondrite \citep{and89} is used. Rate constants for oxidation of individual species are: $(1.17 \pm 0.19) \times 10^{-15}~\mathrm{m^{3} \cdot mol^{-1} \cdot s^{-1}}$ for $\mathrm{Mg^+ + O_3}$ \citep{wha11}, $(6.5 \pm 2.1) \times 10^{-16}~\mathrm{m^{3} \cdot mol^{-1} \cdot s^{-1}}$ for $\mathrm{Si^+ + O_3}$ \citep{gom11}, and $(7.1 \pm 2.3) \times 10^{-16}~\mathrm{m^{3} \cdot mol^{-1} \cdot s^{-1}}$ for $\mathrm{Fe^+ + O_3}$ \citep{rol98}.}
\centering
\begin{tabular}{cccccl}
\hline
Shower & Element & Abundance & Weight & $k$ & Reference \\
 & & (fraction) & $w_i$ & ($\mathrm{m^{3} \cdot mol^{-1} \cdot s^{-1}}$) & \\
\hline
Camelopardalids & Mg & (0.36) & 0.16 & $7.76\times10^{-16}$ & \citet{and89} \\
       & Si & (0.34) & 0.15 & & \\
       & Fe & (0.30) & 0.68 & & \\
\hline
Perseids & Mg & 0.37 & 0.29 & $8.09\times10^{-16}$ & \citet{borovivcka2005elemental} \\
       & Si & 0.52 & 0.55 & & \\
       & Fe & 0.11 & 0.16 & & \\
\hline
Draconids & Mg & 0.42 & 0.18 & $7.79\times10^{-16}$ & \citet{bor07} \\
       & Si & (0.31) &  0.27 & & \\
       & Fe & 0.27 & 0.55 & & \\
\hline
Geminids & Mg & 0.52 &  0.31 & $8.36\times10^{-16}$ & \citet{kas05} \\
       & Si & (0.25) & 0.27 & & \\
       & Fe & 0.23 & 0.42 & & \\
\hline
\end{tabular}
\label{tbl-abundance}
\end{table}

\clearpage

\begin{table}
\caption{Summary of characteristic time measurements and the derived ozone concentrations (in part per million volumes).}
\centering
\begin{tabular}{ccccccc}
\hline
Shower & Radar & Year of obs & $t_c$ & Knee & $[\mathrm{O_3}]$ & Reference \\
 & & & (s) & (km) & ppmv & \\
\hline
Camelopardalids & CMOR & 2014 & $1.6\pm0.1$ & 85 & $3.72\pm0.22$ & \citet{2016Icar..264...48Y} \\
\hline
Perseids & Ond\v{r}ejov & 1991 & $21.1\pm2.5$ & 96 & $2.43\pm0.26$ & \citet{jon95} \\
 & Springhill & 1957--1982 & $34.6\pm4.5$ & 97 & $1.78\pm0.20$ & \citet{jon95} \\
\hline
Draconids & CMOR & 2011 & $2.7\pm0.2$ & 87 & $3.93\pm0.27$ & \citet{ye2013radar} \\
 & CMOR & 2012 & $2.5\pm0.2$ & 89 & $6.07\pm0.45$ & \citet{ye2013unexpected} \\
 & Springhill & 1985 & $13.1\pm0.9$ & 94 & $2.83\pm0.18$ & \citet{sim94} \\
\hline
Geminids & Ond\v{r}ejov & 1958--1991 & $5.0\pm0.4$ & 83 & $1.01\pm0.07$ & \citet{jon95} \\
 & Kharkov & 1958 & $4.6\pm0.3$ & 83 & $1.10\pm0.07$ & \citet{jon95} \\
 & Springhill & 1957--1982 & $13.8\pm0.6$ & 86 & $0.60\pm0.02$ & \citet{jon95} \\
\hline
\end{tabular}
\label{tbl-results}
\end{table}

% Don't change these lines
\bsp	% typesetting comment
\label{lastpage}
\end{document}